\documentclass[sigconf]{acmart}

\usepackage{booktabs} 
\usepackage[portuges,brazilian,english]{babel}
\usepackage[utf8]{inputenc}
\usepackage{balance}

\setcopyright{rightsretained}

\begin{document}

\copyrightyear{2017}
\acmYear{2017}
\acmConference[UCC'17]{10th International Conference on Utility and Cloud Computing}{December 5--8, 2017}{Austin, Texas USA}

\setcopyright{acmlicensed}
\acmPrice{15.00}
\acmDOI{10.1145/3147213.3147230}
\acmISBN{978-1-4503-5149-2/17/12}

\title{Secure and Privacy-Aware Data Dissemination \\ for Cloud-Based Applications}

\fancyhead{} 

\author{Lilia Sampaio}
\affiliation{%
  \institution{Universidade Federal de Campina Grande}
  \city{Campina Grande} 
  \country{Brasil} 
}
\email{liliars@lsd.ufcg.edu.br}

\author{Fábio Silva}
\affiliation{%
  \institution{Universidade Federal de Campina Grande}
  \city{Campina Grande} 
  \country{Brasil} 
}
\email{fabiosilva@lsd.ufcg.edu.br}

\author{Amanda Souza}
\affiliation{%
  \institution{Universidade Federal de Campina Grande}
  \city{Campina Grande} 
  \country{Brasil} 
}
\email{amandasouza@lsd.ufcg.edu.br}

\author{Andrey Brito}
\affiliation{%
  \institution{Universidade Federal de Campina Grande}
  \city{Campina Grande} 
  \country{Brasil} 
}
\email{andrey@lsd.ufcg.edu.br}

\author{Pascal Felber}
\affiliation{%
  \institution{Université de Neuchâtel}
  \city{Neuchâtel} 
  \country{Switzerland} 
}
\email{pascal.felber@unine.ch }

\renewcommand{\shortauthors}{L. Sampaio et al.}

\begin{abstract}
In this paper we propose a data dissemination platform that supports data security and different privacy levels even when the platform and the data are hosted by untrusted infrastructures. The proposed system aims at enabling an application ecosystem that uses off-the-shelf trusted platforms (in this case, Intel SGX), so that users may allow or disallow third parties to access the live data stream with a specific sensitivity-level. Moreover, this approach does not require users to manage the encryption keys directly. Our experiments show that such an approach is indeed practical for medium scale systems, where participants disseminate small volumes of data at a time, such as in smart grids and IoT environments.
\end{abstract}

%
%






\maketitle

\section{Introduction}

The ``digital transformation'' enables an increase in productivity and quality of life through the usage of information technologies. The growing number of data sources combined with analytics techniques that generate actionable information from a large volume of raw data have a strong impact in all aspects of our daily lives. Nevertheless, these opportunities are also coupled with several challenges, especially the need for affordable and scalable infrastructures to hosts data and applications, as well as the mitigation of risks related to the leakage of private data.

On the one hand, the challenge of providing scalable infrastructures has been addressed by the advances in cloud computing. In contrast to the situation in the last couple of decades, where developers of novel applications would have to consider the risk of investing in hardware infrastructures, new applications start today in the cloud, where the cost is proportional to the resources actually used (at the granularity of cents) and the infrastructure can be scaled within minutes. In addition, there are hundreds of cloud providers that offer more than simple computing and storage resources paid by the hour. These providers offer higher level platform services to ease the development of applications. This combination of simplicity and cost efficiency has promoted the cloud as the \emph{de facto} environment where applications are hosted.

On the other hand, cloud providers are an obvious and attractive target for attacks that aim to steal data or compromise applications. There are many reasons that increase the risk of data leakage when using cloud infrastructures~\cite{treacherous}, for example: \emph{(i)} vulnerabilities in the infrastructure may allow attackers to access data outside their VMs or tenants; \emph{(ii)} employers may have access to raw data and use these access to steal data; or \emph{(iii)} cyber-espionage may compromise the confidential data of companies and even governments. At the same time, cloud platform services store increasingly more sensitive information, such as voice snippets, like in AWS Lex\footnote{\url{https://aws.amazon.com/lex/?nc2=h_a1}}, and face images, as in AWS Rekognition\footnote{\url{https://aws.amazon.com/rekognition/?nc1=h_ls}}.

While there are many guidelines for building cloud native applications, if the infrastructure cannot be trusted as are the cases listed above protecting data becomes challenging. Encrypting data at rest, using good encryption keys, and limiting the scope and permissions of the users cannot protect from insider attacks or remote attacks in which the attacker has manage to compromise the physical host. In such scenarios, some approaches, such as homomorphic encryption, are effective even in such cases as the data can be kept encrypted at all times, even during processing. Nevertheless, it is very hard to compile generic functions into an application that uses homomorphic encryption. Finally, the usage of secure co-processors have been considered for decades, but required specialized hardware that was typically not widespread.

More recently, the idea of having secure coprocessors have gained additional traction. It started on the domain of embedded devices with ARM Trustzone, but with SGX has reached common workstations and servers\footnote{Currently, processors with Intel SGX support are the sixth and newer generations processors of the Intel Core family and some recent Intel Xeon processors, such as the E3-1200 family, fifth generation and newer.}. Intel SGX enables code to be executed in a secure enclave in a way that its data is protected even from the operating system. In addition, it supports attestation, where the code running in such secure enclave has its signature validated. After SGX, other mainstream processor manufacturers, such as AMD, have then also proposed similar approaches and with the amount of sensitive data being kept in our machines, the trend is that these hardware technologies will become ubiquitous.

In this paper, we address this problem by exploiting tools that enable the usage of SGX to host communication and processing systems. We than propose a system that combines and extend tools such as Intel SGX, SCONE~\cite{scone2016}, and SCBR~\cite{scbr2016} in a way that enables data producers to be aware of the entities that are going to consume its data (through remote attestation) and even restrict the level of granularity that these entities can consume. Through this combination of features, it is viable to produce an ecosystem of applications in which a data source produces very sensitive data that is repeatedly anonymized or aggregated by trusted entities. Less sensitive versions of the data can then be consumed by less trustworthy (or even untrusted) applications.

The rest of the paper is organized as follows. Section~\ref{sec:background} discusses tools and concepts that are fundamental for the approach presented and a running example that will help illustrate the approach described is Section~\ref{sec:approach}. After that, Section~\ref{sec:eval} presents experiments of the proof-on-concept implementation. The paper is concluded with related work in Section~\ref{sec:related} and some final remarks in Section~\ref{sec:conclusions}.


%
%
%
%
%
%
%
%
%
%
%

\section{Background}
\label{sec:background}

In this section, we provide a brief description of the key concepts to aid the understanding of the context and the components that will be used in our proposed architecture.

\subsection{Intel SGX: Software Guard eXtensions} \label{intelSGX}
\label{sec:sgx}

Securing data in order to guarantee its privacy and integrity is highly desired by end users aiming to protect sensitive information from malicious attacks. Some approaches that attempt to provide this security, specially on cloud environments, lack the ability to protect the application data from software with higher privilege levels such as hypervisors \cite{noHype2011,surveyCloudSec2011}. In this context, Intel's Software Guard eXtensions (SGX) \cite{intelSGX,sgxExplained2016} has emerged, a hardware-based technology that ensures privacy and data integrity, protecting application code even if components such as the OS, hypervisor, etc. are untrusted.

In order to achieve this goal, Intel SGX provides a set of instructions to allow changes in memory access, creating protected areas named enclaves~\cite{mckeen2013innovative}. The enclave page cache (EPC) is where application code and data reside, managed by CPU access control policies, which prevent attacks against its content. Code outside an enclave cannot access enclave memory. However, enclave code can access untrusted memory outside the EPC, being responsible for verifying the integrity of this data.

Intel SGX also offers local and remote attestation features \cite{anati2013innovative}, which can be performed by a third-party to guarantee that an expected piece of software is securely running inside an enclave, on a known SGX-capable platform. Remote attestation, used in this paper, requires asymmetric cryptography, since the verification comes from outside the platform, and a special component, the quoting enclave. This enclave is responsible for creating the Intel Enhanced Privacy ID (EPID) key used for signing attestations to be certified by an EPID backend infrastructure. Only the quoting enclave knows this EPID key, which is connected to the version of the processor's firmware.

Possible usage of Intel SGX \cite{barbosa2016foundations,hoekstra2013using} includes authentication technologies, online financial transactions, logging of user activities and personal information, video conferencing, and many others. Besides these examples on client machines, it is also possible to use SGX to protect backend applications. For instance, VC3 \cite{schuster2015vc3} runs distributed MapReduce computations in the cloud guaranteeing data privacy and ensuring the correctness and completeness of results. VC3 uses SGX to isolate memory regions, and to deploy new protocols that secure distributed MapReduce computations.

\subsection{SCONE: Secure Linux Containers} 
\label{sec:scone}

With the advent of SGX and the growing use of containers for hosting applications, new approaches to handle security and privacy aspects of such structures have emerged. Here, we use SCONE \cite{scone2016}, a Secure CONtainer Environment for Docker that uses SGX to protect given container processes, using SGX protected enclaves. This mechanism offers secure containers together with insecure operational systems, and does that in a transparent way to already existent Docker environments. For this to happen, it is only required that the host machine has a SGX-capable Intel CPU and a Linux SGX kernel driver\footnote{https://01.org/intel-softwareguard-eXtensions (visited: June 05, 2017).} installed.

Amongst the features offered by SCONE, there are (\emph{i}) an asynchronous system call interface to the host OS provided to container processes, allowing them to perform system calls without having to exit threads inside enclaves; (\emph{ii}) support for transparent encryption and authentication of data through a mechanism called \textit{shielding}, ensuring data integrity and confidentiality; (\emph{iii}) no changes to the application code being deployed, since SCONE's special compiler automatically prepares the code to be SGX-compatible; (\emph{iv}) simple Docker integration relying on a secure container image specially built for this purpose.

Besides that, providing a secure container requires a SCONE client extension to enable the creation of configuration files, spawning of such containers and for secure communication with them. During container startup, a configuration file is necessary containing keys for encryption, application arguments and environment variables. Also, the application code must be statically compiled with its library dependencies and the SCONE library. 

In general lines, SCONE provides secure containers maintaining a small Trusted Computing Base (TCB) size, and reducing overheads naturally imposed by SGX enclave transitions, thanks to its asynchronous system calls mechanism and custom kernel module. 

\subsection{Secure Content-Based Routing} 
\label{sec:scbr}

Content-based routing (CBR) is a known paradigm for communication between distributed processes that routes messages based on their content rather than by a specified destination. This allows for more scalability, dynamicity and flexibility, besides removing from the sender application the knowledge of where sent messages will end up. Such publish/subscribe communication mechanism~\cite{pubsub2003, pubsub2012} can be improved by adding an extra security layer to the process, since in this scenario, the router has access to the content of the messages and subscriptions, representing a threat to the data confidentiality and integrity which might be compromised.

Considering this, here we use a Secure Content-Based Routing mechanism~\cite{scbr2016} that relies on the SGX technology previously described to provide a routing engine in an enclave. We add to SCBR features a protocol for exchanging cryptographic keys between both ends of the communication chain, producers and consumers of smart metering data, and the routing engine. As a consequence, because publications and subscriptions are encrypted and signed, the system raises protection levels against malicious attacks that could compromise the data being exchanged. 

\subsection{Python-SGX interpreter}
\label{sec:pythonSGX}


The Intel SGX SDK is a development toolkit available only for C and C++ languages. This means only applications written in these languages can be adapted to run and communicate with enclaves. This presents itself as a limitation for the Intel SGX technology as porting code is an obstacle and may lead to additional bugs. 

Among popular programming languages, Python deserves special attention. This year, Python was considered the Top 1 programming language in the 2017 Programming Languages ranking promoted by IEEE\footnote{http://spectrum.ieee.org/computing/software/the-2017-top-programming-languages}. Very popular softwares are written in Python as well, such as OpenStack, YouTube, DropBox, Instagram and many others. 

Using Intel SDK to implement SGX applications might require extra effort to port existing code, or even creating new pieces of software. For this purpose, SCONE provides a modified C compiler, based on the \textit{libmusl}\footnote{https://www.musl-libc.org/} library. This compiler, named \textit{sgxmusl-gcc}\footnote{https://sconedocs.github.io}, automatically generates the object code to be executed inside SGX enclaves, making it easier to have hardware protected applications ready to run. However, the \textit{sgxmusl-gcc} compiler is obviously restricted to C code, and possibly with GCC supported languages, such as Fortran, through the \textit{libgfortran} library\footnote{https://gcc.gnu.org/wiki/GFortran}.

In the light of this, and the increasing use of the Python language mentioned before, enabling Python code to run in SGX becomes attractive. We then leverage the \textit{sgxmusl-gcc} to produce a modified Python interpreter. Our Python interpreter is compiled with the \textit{sgxmusl-gcc} and extended to interpret and attest Python code inside SGX enclaves. All things considered, this approach increases the range of applications that can be executed using SGX
as well as the number of developers capable of leveraging the technology. 

However, the \textit{sgxmusl-gcc} compiler has a few limitations. One of the major limitations is the fact that dynamic linking of libraries is not allowed. 
All the system libraries, such as \textit{openssl}\footnote{https://www.openssl.org/} and \textit{ncurses}\footnote{https://en.wikipedia.org/wiki/Ncurses}, together with the native Python modules required by the user's application, should be statically linked upon Python-SGX building.

Unfortunately, it is not possible to include all native Python modules at once. Static linking requires the code from all the libraries to be included in the binary file, causing a large memory overhead upon execution. Adding extra code also introduces the risk for bugs in the generated code. In practice, limiting imports is not necessarily critical as most applications, even highly complex ones, are unlikely to use too many libraries. This observation is specially true when considering the microservice approach, where functionally is well divided into a large number of services.


When it comes to external libraries, there is a level of complexity added when they are not pure Python. By default, it is not possible to interpret application code that requires such modules. However, we managed to support some important ones as the PyCrypto\footnote{https://pypi.python.org/pypi/pycrypto} library. This Python cryptography toolkit provides a stable and trustworthy base for writing Python code that requires cryptographic functions, such as the AES-CTR encryption mode used in this paper. To make the link possible, we had to introduce PyCrypto as a native Python module. To do this, the PyCrypto library had to be modified to be included in the Python-SGX source tree, and then able to be linked as the rest of the native libraries.

In addition, we also use Python-SGX to interpret the code for some of the components explained in Section \ref{sec:components}, and we attest them in a way that guarantees the code is the one expected by the developer. We do this by introducing in the interpreter code checks for the SHA-256 hash of the application code. The hash of the code to be executed is calculated before the start of the interpretation and checked against the hash provided during the SCONE attestation process of the Python binary itself. In the big picture, Python-SGX is considered trusted because it is previously attested by SCONE and this trust is extended to the Python code executed over it.

In a summary, considering the above limitations, we managed to make possible for complex applications written in Python to be interpreted by our Python-SGX, and therefore, securely executed inside SGX enclaves in a transparent way. 








\subsection{Application example: a smart metering infrastructure}
\label{sec:example}

For detailing the approach proposed in this paper we consider a smart metering use case. The motivation for such an application scenario is that the availability of detailed power consumption information enables analytics that can reduce power consumption by detecting anomalies and undesired configurations, recommending actions that will result in more efficient usage of the electricity. 

As an example, collecting measurements at each second may enable the identification of individual appliances running in a consumer's unit. This is known as Non-Intrusive Load Monitoring (NIALM)~\cite{nilm}. With this information, customized recommendations can lead consumers to save considerable amounts of power~\cite{holygrail}. Nevertheless, even without disaggregation, the usage of detailed metering for home energy management systems~\cite{hems} has proved its value in practice.

On the negative side, providing detailed power consumption information reveals much more than it may seem at the first glance. Previous research has shown that even details on the multimedia content in users' TVs may be detected through detailed data~\cite{multimedia}. It then becomes clear that even less information can reveal much about the habits of individuals of a residence.

In a summary, having detailed power consumption is clearly useful. Power utilities can use the data to better plan power generation and to understand and, therefore, influence consumers. Consumers may benefit from analytics approaches being executed over its data. However, even with clear benefits, the data should not be trusted to any application. In addition, not every customer will want to share his data. Consequently, a system that enables users to have better control over who access the data and reduces the risks of leakage can be a seed for sophisticated privacy-aware applications not only in smart grid infrastructures, but also in other smart cities and IoT application domains.

\section{Privacy-aware data dissemination}
\label{sec:approach}

This section describes our data dissemination platform. It begins by describing the basic components and then continues to describe a simple publication workflow. The description ends by detailing how the example introduced in Section~\ref{sec:background} can be improved based on the platform.

\subsection{Components}
\label{sec:components}

\subsubsection{Smart Meter}

Smart meters are a key component for smart grids. Such devices are responsible for collecting energy data of households, buildings, and other environments, enabling customers to reduce electricity costs by wisely monitoring their energy consumption. A smart meter can read these fine-grained measurements at specified time intervals, and communicate this information to a utility provider. For future generations smart energy systems, it is expected that not only the utility will consume this data, but also third parties will offer applications that monitor consumption and recommend efficiency actions, as discussed in Section~\ref{sec:background}.

In our scenario, we consider the Smart Meter component a device that is able to directly or indirectly send data to a remote server. In practice, because of regulatory or cost constraints this is typically done indirectly, meters send data to gateways and these send the data to a processing system at the utility. 
Nevertheless, because of this indirect communication, there is much flexibility in the implementation of the communication. In our proof-of-concept we consider different meters that can be accessed through a wireless or cabled network. We then consider that energy consumption data is collected by a Metering Data Collector (MDC) component that may be specific to an equipment model or brand and then the measurements are forwarded to other systems.

\subsubsection{Metering Data Collector}

This component is responsible for collecting energy consumption data from the smart meter device. The MDC application connects to the device via a TCP/IP network and retrieves new data every second. The communication protocol may be specific to that device. Because the next component, named \emph{Dispatcher}, is untrusted, encryption is needed to guarantee some security. In our use case, this untrusted component is the Dispatcher, detailed below. This encryption is currently implemented using the AES-CTR encryption mode (possibly a rotating key). 

In addition, because the sole purpose of the key is to protect from the untrusted dispatcher, there are two ways this key can be created and managed: (\emph{i}) the key is negotiated with the SCBR during the attestation process (described below); (\emph{ii}) the key will be generated by the data source and shared with all trusted participants in the system. As we will detail later, the first approach introduces higher load to the SCBR, reducing its scalability. In contrast, the second approach requires some periodic rotation of keys.


Because the MDC sees raw data it needs to be trusted. This trust can be gained through certification and sealing (as the meters typically are) or through the usage of a trusted execution environment. In our case we consider the second. Therefore, the MDC is executed in a special Python interpreter that was generated with SCONE (as detailed in Sections~\ref{sec:scone}~and~\ref{sec:pythonSGX}). By executing the MDC inside an SGX enclave, we are able to validate this code before execution, ensuring that only versions with the expected signatures will be executed.

\subsubsection{Dispatcher}


In our scenario, the Dispatcher works as a gateway, passing along the measurements received from the MDC application to the message bus. Because the MDC may be limited in functionality, the usage of the Dispatcher enables further flexibility in the setup of the rest of the system. Furthermore, because the Dispatcher does not need to be trusted, it has many more implementation and deployment options. As an example, with an untrusted dispatcher, it is trivial to change the message bus if SCBR guarantees are not needed.

In our specific implementation, the communication with the bus requires ZeroMQ\footnote{http://zeromq.org/} connections and the Python-SGX interpreter has limited support for importing Python modules. As it is likely to occur in practice, by using an indirection level we decouple a trusted component, the MDC, from the communication protocol used by the data infrastructure running in the cloud. This decoupling eliminates the need of reimplementing the bus's communication protocol, which would possibly add more complexity to the proposed platform and its usage.

Thus, the Dispatcher simply implements a layer of communication with the bus via ZeroMQ and communicates with the MDC through simple sockets. Finally, the received encrypted measurements are sent to the Secure Content-Based Routing (SCBR) component.

\subsubsection{Secure Content-Based Routing}

The Secure Content-Based Routing component follows the publish–subscribe paradigm~\cite{pubsub2003}, in which senders of messages, named publishers, do not address messages explicitly, but rather categorize such messages regardless of which receivers, the subscribers, will be receiving them. From the subscribers' perspective, subscribers express interest in one or more type of messages and receive the ones they are interested in, regardless of their publishers.

Differently to other regular publish-subscribe middleware, SCBR has a mode in which only the publisher can submit subscriptions to its publications. We use this mode so that subscribers have to communicate with the publishers at the beginning. During this initial communication, the publisher will attest the candidate consumer, and if passed, it can handle encryption keys.

The SCBR bus securely routes messages between publishers and subscribers, as detailed in Section~\ref{sec:scbr}. Its security and privacy-awareness are consequence of the fact that the routing decisions are taken inside SGX. In our use case, we consider that sensitive information will carry its sensitivity level (in the Pub/Sub topic). 


Depending on the choice on the encryption approach in the MDC, as discussed above, there are two choices: (\emph{i}) if the MDC encryption key was negotiated with the SCBR, SCBR would decrypt the data and this data would be disseminated unencrypted; (\emph{ii}) if the encryption key is negotiated with the trusted parties, the sensitive information (e.g., the measurements is kept encrypted even within the SCBR enclave), this could be useful for connecting systems that store information, even if the systems themselves cannot read it.



Sent messages follow a specific header format, containing the message type, or privacy level, and its encryption mode which can also be \textit{plain-text}. SCBR allows for configuring the security level of its core, which can be set to use SGX or not. The bus uses the Intel SDK tool in its implementation and, by enabling SGX, executes the routing engine inside an enclave. From the client side, if the SGX mode is required, it can be checked as the attestation process only works if the SGX enclaves are in use.

\subsubsection{Attestor}

In the proposed platform, in order for consumers to receive data from SCBR, besides being registered in the bus, they need to be considered trusted and therefore, attested. With this approach, the consumer knows how to decrypt the encrypted measurements, and can have access to the published information.

The attestation process here follows the remote attestation protocol specified by SGX, as explained in Section~\ref{intelSGX}, and uses the Intel Attestation Service (IAS). The Attestor is then responsible for mediating the attestation process of the consumers by the IAS, and then, the encryption/decryption key exchange during attestation. The process is further discussed in Section \ref{archFlow}.

\subsubsection{Aggregator}

Aggregating individual measurements to produce full energy consumption reports and its respective billing is an important feature desired by utility providers in a smart metering scenario. Here, the Aggregator component serves this purpose and aggregates measurements generating new aggregated energy data. Time intervals may vary between minutes, hours, or months, but every message received by the Aggregator is an individual raw measurement as previously published by the Smart Meters. 

During initialization, the Aggregator is attested by the IAS through the Attestor, and is then able to decrypt the received messages. After the key exchange, this component is also able to encrypt the aggregated data to be published again into the SCBR bus and later consumed by final consumers. In our implementation, this piece of code was written using the Intel SDK.

\subsubsection{Final Consumers}

There can be many final consumers, which are able to register to the SCBR bus and express interest in a certain type of messages. For the registration, they need to contact publishers and then will be attested by the IAS through the Attestor. As a consequence, they will receive the key to decrypt published messages. This piece of code is also developed using the Intel SDK tool. An alternative is that the consumer requests public data. In this case, the publisher would simply register it without actual attestation.

\subsection{Architecture Workflow} 
\label{archFlow}

\begin{figure}
\begin{center}
\includegraphics[width=.450\textwidth]{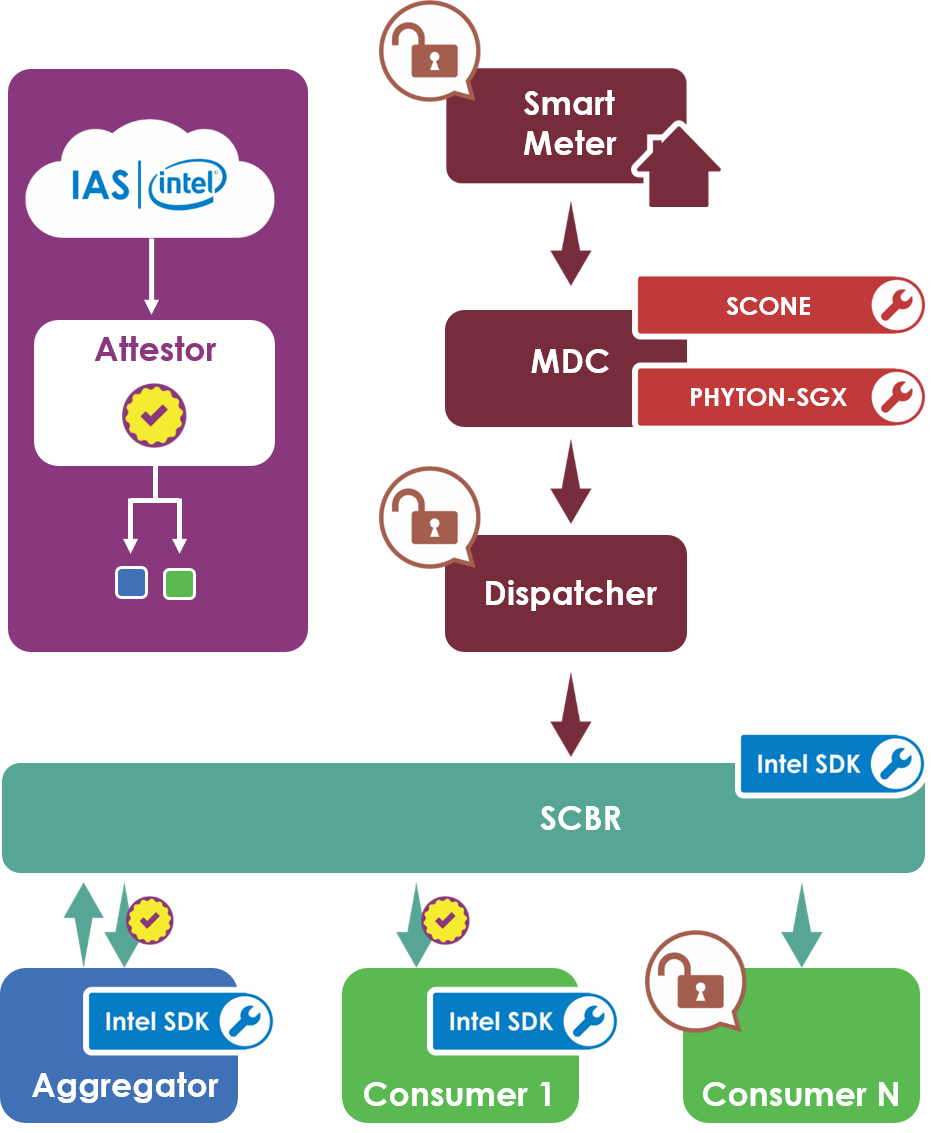}
\end{center}
\caption{Architecture of data dissemination platform for smart metering infrastructures.} 
\label{fig:arch}
\end{figure}


As seen in Figure \ref{fig:arch}, the flow starts with measurements being recorded by Smart Meters. These data is then sent to the MDC application, which is interpreted by the Python-SGX interpreter. Python-SGX is attested by SCONE, and afterwards, can attest the MDC component and guarantee that the SHA-256 hash of the application code matches the hash provided during the SCONE attestation process.

The MDC then collects measurements every second from the Smart Meter, via a secure HTTPS channel, and encrypts them using the AES-CTR encryption mode. The key used for encryption is generated from an Initialization Vector (IV) and the respective decryption key is handed in during the consumer's attestation process. The MDC then sends the encrypted measurements to the Dispatcher via TCP sockets. To complete the publication flow, the Dispatcher communicates via a ZeroMQ connection with the SCBR bus, and publishes the encrypted measurement according to its privacy level.

From the same figure we can also see that the bottom half of the SCBR component represents the consumers interested in the messages published by the Dispatcher. The figure illustrates two types of consumers, Aggregators and Final Consumers, as described in Section \ref{sec:components}. In our scenario, we have one aggregator and a number of final consumers. All of them should first register to the bus through the producers, and declare which type of messages they are interested in. Upon registration, such consumers are attested by the IAS through the Attestor component. This process is indicated in Figure \ref{fig:arch} by the yellow ticks over the arrows connecting the consumers to the bus. When the attestation process is completed, the consumers will receive the key able to decrypt the published encrypted messages. This shared key is sent encrypted by a symmetric key also negotiated during the attestation.

From this point on, consumers are able to decrypt the messages received from the SCBR bus. By definition, the aggregator receives the encrypted raw measurements and aggregates them according to specific time frames previously defined. These frames may vary between seconds, hours, months, and so on. After the data is aggregated, it can be encrypted or not, and published again to the SCBR bus by the aggregator itself, which in this scenario also works as a publisher. Upon publication, the privacy level of the information is defined, and the final consumers receive them accordingly.








\begin{figure*}
\includegraphics[width=.9\textwidth]{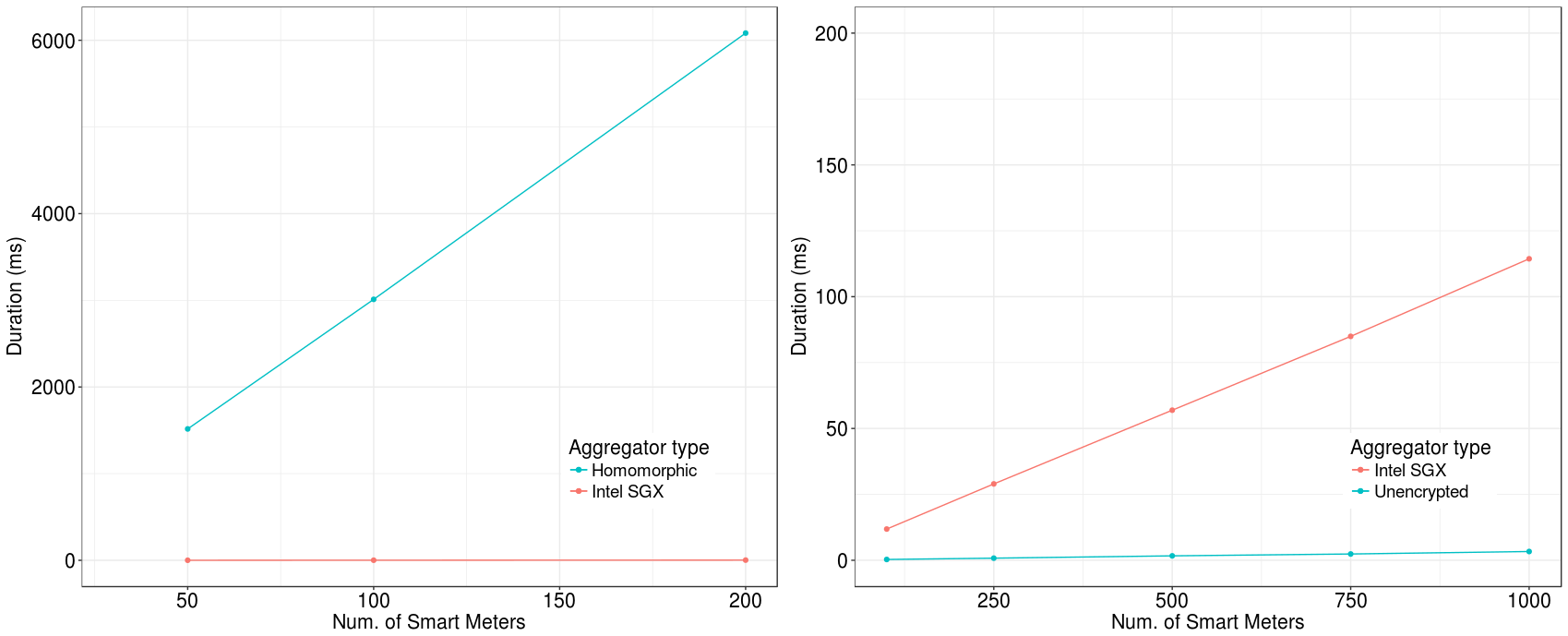}
\caption{Example of HE vs. SGX.}
\label{fig:hecomparison}
\end{figure*}

\subsection{Securely aggregating measurements}
\label{sec:exempleimproved}

As discussed in Section~\ref{sec:background}, we envisage a scenario where a meter will collect detailed information that is valuable for many uses. For example, a user may be interested in receiving customized recommendations on how to reduce consumption. For that, it opts in to an application that request access to its second-level data. Another user, concerned about privacy, does not take part in such application and allows only aggregated measurements (e.g., daily) to be accessible.

For our running example, we classify the impact level about criticality and sensitivity of aggregate data. We adjusted the FIPS 199~\cite{NIST2004} model as an assessment criteria in the current example. The security model for data aggregation classifies potential impact in three levels: Low, Moderate or High. These levels can be interpreted as follows:

\begin{enumerate}
\item High impact characterizes private data (e.g., an individual data collection from a particular smart meter), and the information must only be available to an aggregation system, protected by enclaves, or to applications explicitly trusted by producers (the smart meter owners). The risk of exposure or data violation would be unacceptable, resulting in severe or catastrophic impact or noncompliance with legal requirements and loss of customer confidence. 

\item Moderated impact defines protected data (e.g., as a set of local data aggregation from a collection of smart meters), and the measurements should only be available to power supplies. The risk of exposure or data violation would be marginally acceptable, causing certain impact or reputation losses on normal activity, with adverse effects on organizational operations, assets or individuals.

\item Low impact identifies public data (e.g., as a set of aggregated data from regional smart meters), and the information may be available freely. The risk of exposure or data violation would be acceptable for the energy company, resulting in minimal impact on normal activity. 
\end{enumerate}

Thus, in our example we consider that meters high-frequency measurements are published by the meters as \emph{high impact}. These data are subscribed to by the Aggregator and by the Consumer $1$, both have been explicitly trusted by the Smart Meter owner. The Aggregator computes hourly averages from the consumer measurements (e.g., for billing purposes, compatible with hour-of-the-day tariffs) and regional hourly averages (e.g., for public viewing). These two aggregations are published as \emph{moderate} and \emph{low-impact}, respectively. Consumer $N$ subscribes to low-impact measurements and receives these publications even though it cannot be attested, as it is not running in an enclave.





\section{Evaluation}
\label{sec:eval}

In this section we discuss the experiments that validate the proposed architecture. All experiments were executed in workstations with an Intel Core SGX-enabled $i7$-6700 processor and $8\ GB$ of RAM running Ubuntu Linux 16.04 Xenial.

In the first set of experiments, we compare a simple aggregation implemented with homomorphic encryption, which would also enable trusted data processing in untrusted infrastructures, with SGX (but without any other components of the proposed architecture), and in pure C, without SGX. The implementation used in the homomorphic encryption aggregation is based on the scheme proposed by Busom et al~\cite{busom2016efficient}. The goal is to illustrate how the raw SGX performance compares to a traditional approach that would enable comparable benefits for running protected data analysis in cloud environments, and to an approach that offers no security, but also no overhead.

Because of the huge discrepancies in the overhead two sets of tests were executed. The first compares homomorphic encryption and Intel SGX and the second compares Intel SGX with the pure C implementation. The results of the first tests are depicted in Figure~\ref{fig:hecomparison}. Each point in the figure is calculated from 10 experiment runs. The tests are split into two parts: on the left-hand side, it considered that $10$, $50$, $100$ and $200$ meters (or measurements) would be aggregated together to mask individual reads; on the right-hand side, the experiments consider $200$, $400$, $800$, and $1000$ measurements.

The remaining experiments consider the proposed platform.  The smart meters were simulated from regular processes that would connect to the MDC component (see Section~\ref{sec:components} for details). The experiments process up to 1 million measurements. Typically, two curves or box plots are shown: one, described as \emph{SGX}, illustrates executions where Intel SGX usage was enabled; the other, described as \emph{regular} considered executions in which SGX is disabled in the trusted components. In both executions, SGX and regular, the components use AES-CTR encryption for the measurement data sent through the bus. For each scenario that considers a specific data rate, the configurations were repeated $60$ times. The CPU measurements consider the usage of the SCBR processes.

\begin{figure}
\includegraphics[height=5.5cm]{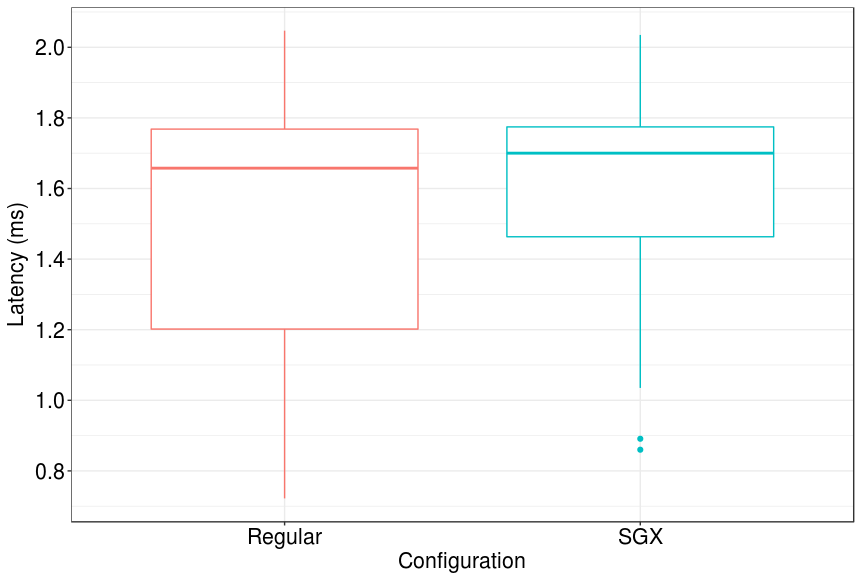}
\caption{Latency for an isolated publication.}
\label{fig:onemeasurement}
\end{figure}

In the experiment depicted in Figure~\ref{fig:onemeasurement} the system was under very light load. A single measurement was published each second. In this scenario, it is possible to see that the latency for publishing a single measurement, passing through a message bus in the same physical host is not statistically different between the two configurations. We can also see that the higher latency value is similar for both configurations, meaning the worst case scenario when using SGX also happens without it.

\begin{figure}
\includegraphics[height=5.5cm]{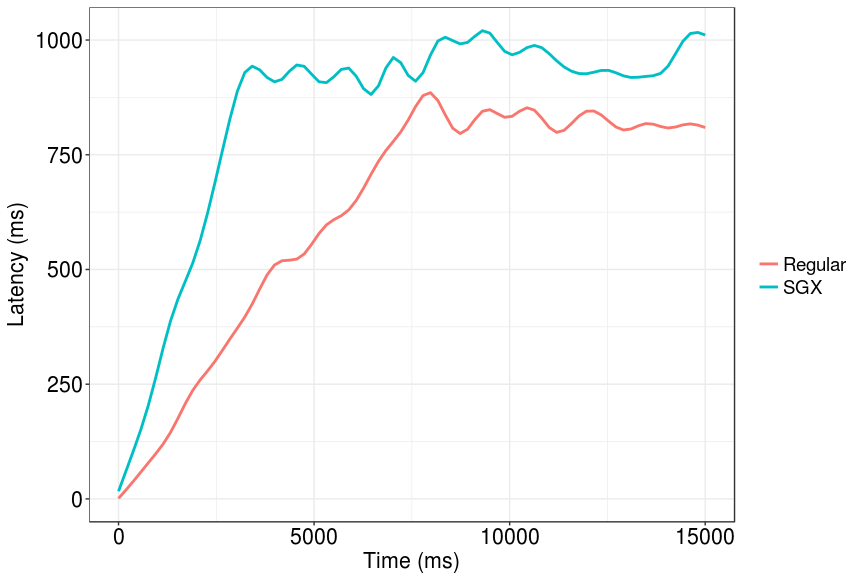}
\caption{Latency for a burst of 1 million publications.}
\label{fig:latencyburst}
\end{figure}

Next, because experiments considering one isolated measurement might not be representative for more complex execution scenarios, we analyzed the system's behavior under a heavier load, processing a burst of 1 million publications from a single producer. Figure \ref{fig:latencyburst} depicts the first 15 seconds of execution. The latency for publishing the measurement passing through SCBR shows an increasing behavior in the beginning of the execution, which means the message bus receives as many measurements as possible until its internal queues are full. From that moment on, we see only a small variance in latency, between 900 and 1000 milliseconds. This can be explained as a type of back-pressure mechanism, which in our case means the bus causes the transmitting device to hold off on sending data packets until the bus's bottleneck has been eliminated. We can see that such condition happens in both configurations, only in different moments, happening a slightly earlier when using SGX. After the bus' saturation occurs, the latency for both cases is similar. 

\begin{figure}
\includegraphics[height=6cm]{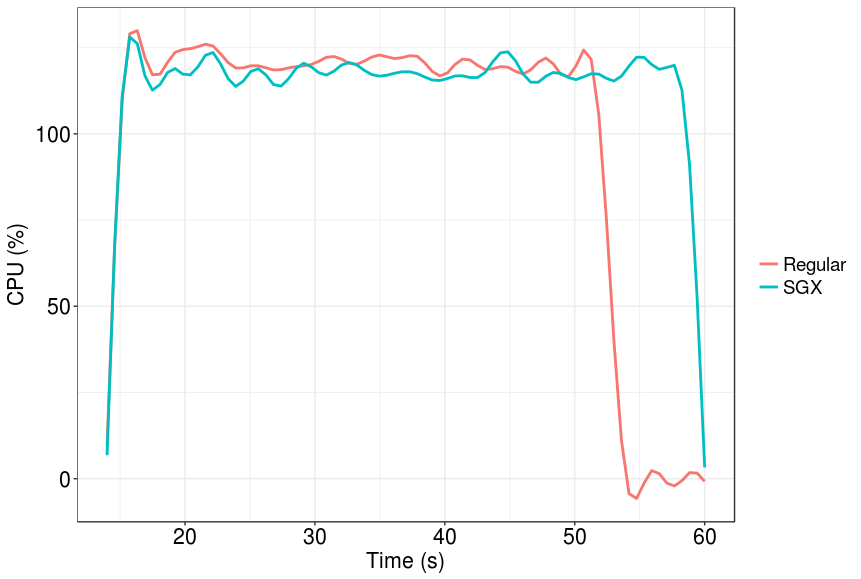}
\caption{CPU usage for a burst of 1 million publications.}
\label{fig:cpuburst}
\end{figure}









Considering the same heavy load scenario of processing a burst of 1 million publications, Figure \ref{fig:cpuburst} shows that the CPU usage maintains the mean identified for the higher power consumption publication rate, which is around $20000\ measurements/s$, for both configurations (as will be detailed shortly). We can also see that in a regular scenario, the time to process all the publications is around 53 seconds, which is smaller then when SGX is being used. Seven seconds later, in case SGX is enabled, the processes are finished. 





\begin{figure}
\includegraphics[width=.5\textwidth]{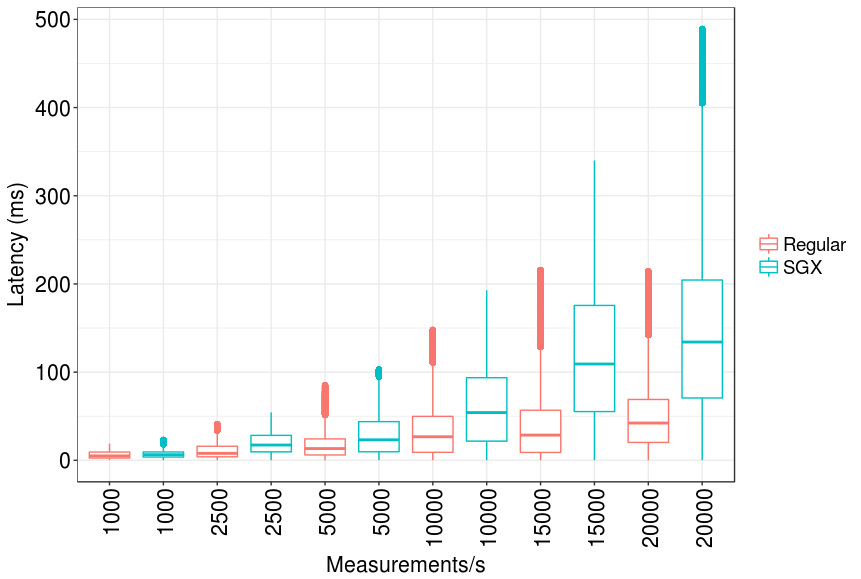}
\caption{Latency for specific measurements rates.}
\label{fig:latency_nmeasurements}
\end{figure}

\begin{figure}
\includegraphics[width=.5\textwidth]{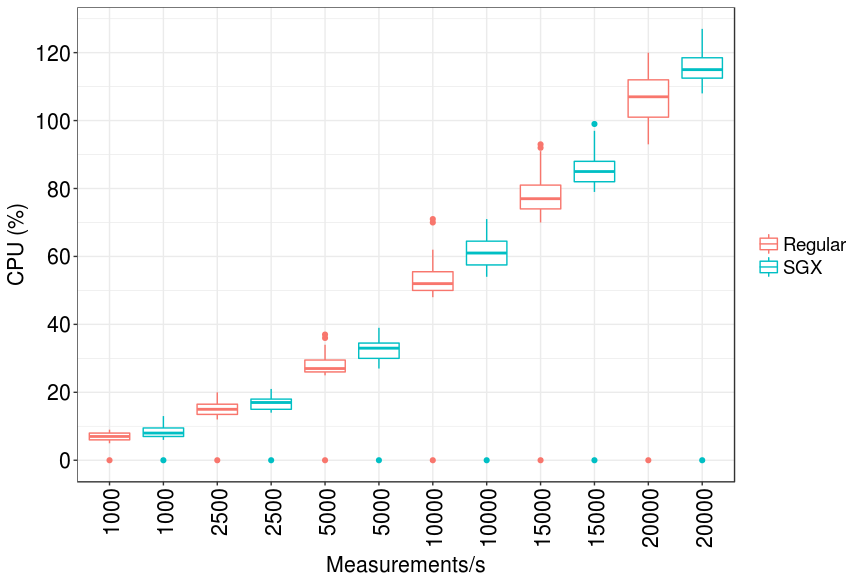}
\caption{CPU usage for specific measurements rates.}
\label{fig:cpu_nmeasurements}
\end{figure}

Experiments considering a variation of publication rates can be seen in Figures~\ref{fig:latency_nmeasurements}~and~\ref{fig:cpu_nmeasurements}. Considered rates were $1000$, $2500$, $5000$, $10000$, $15000$ and $20000$ measurements per second, being the last one the guaranteed number of publications processed in a second without accumulated delays. In Figure \ref{fig:latency_nmeasurements}, we see that only from $15000\ measurements/s$ the mean latency for both configurations starts to differ considerably. For smaller rates, differences are barely noticeable. 







Figure \ref{fig:cpu_nmeasurements} then depicts CPU consumption of the message bus. For both configurations, the usage increases as the publication rates increase. From the results we can see that there is no significantly difference between regular scenarios and when using SGX.















\begin{figure*}
\includegraphics[width=\textwidth]{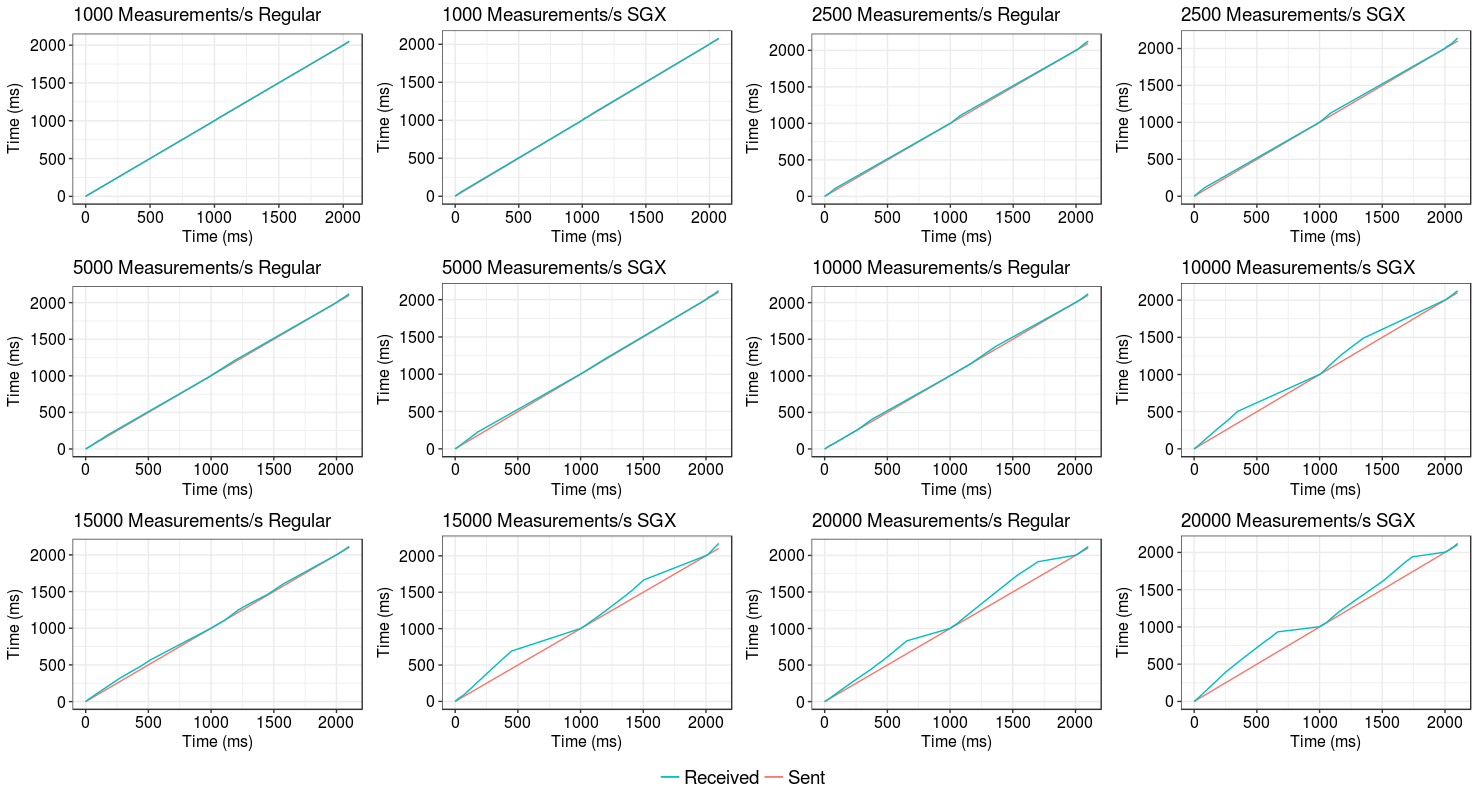}
\caption{Detailed latency considering a variation of measurements rates.}
\label{fig:latency_all}
\end{figure*}

Our last experiment depicts the behavior of the system in the presence of periodic intense bursts of events. Each sub-figure of Figure~\ref{fig:latency_all} depicts one step in the progression from executing experiments with $1000\ measurements/s$, up to $20000\ measurements/s$, in each step we can see how the actual delay deviates from the ideal delay. For example, the ideal curve is depicted in red and is a straight line with an 45 degree angle. For some publications rates, the actual latency deviates from this perfect behavior.

Figure \ref{fig:latency_all} depicts the executions in a 2-second time frame. This value was chosen because it depicts two cycles, hinting the recurring behavior, while not making the figure unreadable. We can see that for rates up to $5000\ measurements/s$, there is no significant visible latency deviation, the lines almost overlap each other. From $10000\ measurements/s$, we can identify a small latency when using SGX, and more clearly for both configurations when considering $20000\ measurements/s$. 







\section{Related Work}
\label{sec:related}

There is a number of prior works focusing on privacy-protecting data dissemination in untrusted environments~\cite{Chhabra:CC,Werner2015,Jiang2012}. We further reviewed secure data dissemination solutions to understand its contributions in the context of cloud computing, focused into two fronts: privacy-preserving data and privacy-assured models for smart metering.

In order to find a solution to address the weaknesses on data exchange between cloud users and providers, the paper by Komnunos and Junejo~\cite{Komninos2015} proposes an encryption scheme which uses a cipher-text policy to anonymize attributes behind the implementation of brokering services and, therefore, protects the data against privacy attacks on cloud environments. 
The work does not consider topics as trust and security issues from a user perspective and its effects when defining data privacy policies.


The authors in~\cite{Itani2009} propose a Privacy-as-a-Service model, defined as a set of security protocols 
to provide a trusted environment that protects confidential data from unauthorized access in cloud infrastructures. 
The authors argue that cryptographic co-processors, used in their solution, demand high resource requirements, and this might be an expensive cost for the privacy gain they offer.

In~\cite{Garcia:2010}, the authors presented a trustworthy system relying on an additive homomorphic encryption to ensure the privacy in smart metering infrastructure. The encryption scheme implements a modular addition operation by using shared keys and provides exchange random measurements among smart meters and utility suppliers. The protocols proposed could be archived using inexpensive smart cards. However, the aggregator must be previously attested to guarantee a trusted service.

Due to the high calculation cost of homomorphic encryption in data aggregation models, the solution in~\cite{Ushdia2013} proposes to perform data mining from secured data avoiding the need of users’ secret key to access confidential aggregation of cloud users. The Privacy Preserving Data Mining (PPDM) technique satisfies the privacy of information stored in the cloud by the granularity level of the aggregation measurements in accordance with the level of secret key carried out of the service providers. 

Another proposal~\cite{Silva:2017} makes use of Intel SGX technology to provide a simple solution for privacy-preserving smart metering system. The authors have suggested a model designed by trusted smart meter devices exchanging consumer's measurements with the trusted aggregator, performed inside an SGX enclave, among a secure channel established during the attestation process.

\section{Conclusions}
\label{sec:conclusions}

In this paper, we have proposed a system that enables a data source that produces sensitive data to control the usage of this data by third parties. One application example is an Advanced Metering Infrastructure, where sources produce detailed metering that may reveal privacy-sensitive information such as consumer habits or, in worst cases, even detailed activities (e.g., TV preferences). In this example, the original data is consumed only by applications and services with higher levels of trust. Among these trusted services, there could be anonymizers and aggregators that reduce the sensitivity of the data to make it consumable by less trusted services. In contrast to other approaches such as homomorphic encryption, our approach has a much lower overhead and can be more simply applied. We demonstrated its usage and feasibility through a set of experiments.
As limitations, the approach depends on the usage of a specific hardware-supported extension, Intel SGX, but which is increasingly common in off-the-shelf machines. 
In addition, Intel also offers SGX virtualization by providing a modified KVM\footnote{https://github.com/01org/kvm-sgx/}, capable of exposing SGX features of hosts to guest VMs, letting users have access to a preallocated EPC memory size. Virtualization support considerably reduces the obstacles of putting SGX support for cloud-based VMs. We have successfully configured this upstream KVM in OpenStack, the most-used open-source cloud management platform.
Finally, our approach also does not protect against software bugs that could reveal or leak data. 

\begin{acks}
This research was partially funded by EU-BRA SecureCloud project (EC, MCTIC/RNP, and SERI, 3rd Coordinated Call, H2020-ICT-2015
Grant agreement no. 690111) and by CNPq, Brazil.
\end{acks}

\bibliographystyle{ACM-Reference-Format}
\balance 

\end{document}